\documentstyle[aps,twocolumn,epsfig,rotating]{revtex}

\begin{document}

\draft \preprint{TRIUMF/McMaster U.}
\twocolumn[\hsize\textwidth\columnwidth\hsize\csname    
@twocolumnfalse\endcsname                               
\vspace{15mm}
\begin{title} {\bf Dynamics of Multidimensional Secession}

\end{title}


\author{Arne Soulier and Tim Halpin-Healy}

\address{Physics Department, Barnard College, Columbia University, NY NY 
10027-6598} \vspace{-5truemm}

\vspace{-4truemm}

\date{September 1, 2002}

\maketitle

\begin{abstract}

  
  We explore a generalized Seceder Model with variable size selection groups and higher dimensional genotypes, 
uncovering its 
well-defined mean-field limiting behavior. Mapping to a discrete, deterministic version, we pin down the upper 
critical size of 
the multiplet selection group, characterize all relevant dynamically stable fixed points, and provide a complete 
analytical 
description of its self-similar hierarchy of multiple branch solutions.

\end{abstract} 



\vskip2pc]                                              

\narrowtext

Dynamical phenomena in which an initially homogeneous population of weakly interacting individual agents can 
disperse, aggregate and 
form
clusters arises in many different physical, biological,
and sociological contexts. Condensation and droplet formation~\cite{drops} is, of course, a well-known example in 
physics; 
galaxy 
formation and clustering ~\cite{astro}, another.  In traffic patterns ~\cite{Traf2},
the real-space jams that plague highway driving are, for some, a daily reminder
of such intrinsic tendencies in correlated systems far-from-equilibrium. The formation of swarms and herds in 
zoology
~\cite{Zoo1}, or the flocking of birds ~\cite{Bird1,Bird2}, provide additional illustrations.  In these cases, 
particularly, 
joining the group yields advantages over standing
out alone, be it by better exploration of food resources,
protection from predators, or easing the aerodynamic flow in flight. 
Nevertheless, sometimes, as in fashion
trends and similar social (or even financial) settings, standing apart from the crowd
can also be a seed for the formation of new groups, splitting
off the mainstream, though maybe becoming the mainstream
themselves later on.
In these instances, steady-state multiple groups can be the norm. Such matters are manifest in 
recent, though now 
classic
implementations of Arthur's variant of the {\it El Farol} Bar problem ~\cite{ElFarol3}, as for example, discussed 
by Zhang and 
Challet
~\cite{ElFarol2}, where a multitude of competing agents, armed with limited memory strategies, compete via
statistical 
Sisyphian dynamics to be in the minority group.   Interestingly, with stochasticity introduced to the 
decision-making 
process, Johnson and coworkers
~\cite{ElFarol1}  uncovered a tendency towards {\it self-organized segregation} within such evolutionary minority 
games.  
Subsequently, Hod \& Nakar ~\cite{Hod} discovered a dynamical phase transition in this setting, between 2-group 
segregation and 
single group clustering, driven by the economic cost-benefit ratio defined in the model. In biological systems,
clustering can appear on multiple scales~\cite{Sca1}, with aggregation a consequence of dynamic correlations, 
whether they be hidden or 
explicit.
Even so, the complexity that arises, for example, in statistical 
models of evolution ~\cite{evol}, resulting in the formation of species is not, per se,
self-evident via the direct interplay of mutation
and selection- the system can dynamically bring itself to a critical state. 
The Seceder Model~\cite{Sec1} was introduced, initially, 
to demonstrate that an interative mechanism favoring {\it individuality}
cannot only create distinct
groups, but also yields a rich diversity of cluster-forming
dynamics.
The essential tack was to give a small advantage to individuals that
distinguish themselves from others. This is not unnatural, since
in epidemics, for example, genetic differences can enhance long-term survival probabilities. Likewise, for players in a minority
game, distinctness may be the advantageous property ~\cite{ElFarol3,ElFarol2,ElFarol1,Hod}.  In this Letter, we 
consider the Seceder 
Model in its broadest sense. Its description is 
simple enough:
within a population of individuals, each described by a genotype variable, choose a subset from the population and 
calculate 
its average.
From within this selection multiplet, the individual most distant from the mean is the parent to be.
Create an offspring by taking this parent's value plus a small
 uniform deviate. 
Finally, replace a randomly chosen member of the population by this new
offspring.
The process is then iterated through many generational time-steps.

Despite the complexity of its segregative dynamics, the Seceder Model may be amenable to 
traditional methodological approaches.  For example, the nontrivial scaling exhibited by the Seceder 
envelope, as well as the interesting 
time evolution of the group number could be thought of as fluctuation-dominated 
non-classical behavior.  In this spirit, it is natural to consider the dimensionality 
dependences inherent in the model, with the expectation of finding a
simpler, mean-field or classical nonequilibrium dynamics within some sector of this larger
parameter space.  Some wisdom in this regard may be had from the Bak-Sneppen model of punctuated 
evolution~\cite{Bak},
wherein a similar, innocuously trim update algorithm engenders an
extraordinarily rich spatiotemporal dynamics.
Even so, a mean-field limit of this model was
subsequently engineered~\cite{Fluv,BD} and further explored~\cite{MM}, simplified scaling retrieved by introducing 
system-wide correlations to the interactions.  Here, for the Seceder
Model, we're motivated by similar goals. Clearly, the number of
parameters is restricted- we have the population size $N,$ which is understood 
to diverge in the thermodynamic limit.  There is also the size $m$ of the
multiplet selection group; finally, the dimensionality $d$ of the genotype variable,
which determines the nature of the base-space through which the population groups
mark their trajectories, provides an additional degree of freedom.

We begin by enlarging the selection group from which the parent is chosen.
Naively, we'd expect the limit $m\rightarrow N,$ which introduces increasing cross-correlation within the
society, to elicit eventually, a {\it mean-field} type of behavior, if only in the extreme
case when $m=N,$ when we're averaging over the entire population using
the societal mean to determine the most distant, reproduced individual. 
Indeed, this is the case. The surprise, however, comes with
the abruptness of the transition. There is, already, a marked change of behavior as 
we switch from a triplet $(m=3)$ to a quartet $(m=4)$ selection group. 
In Figure 1, we show single runs of the $d=1$ Seceder Model using multiplet groups 
$m=3-8,$ within an essentially infinite population, $N=512.$  For $m=3,$ we
have trademark Seceder demeanor, with self-similar branching
characterized by three dominant, but fluctuating arms, centered about the 
origin, with ample small-scale stochastic structure associated with the transient appearance of
variously short-lived subbranches. Rather than a gradual transition, we find for $m=4$ that the 
typical stable configuration {\it suddenly} involves two groups, not three.  In addition, these
two branches exhibit only the most modest sorts of fluctuations, as is evident
from the figure. Increasing the selection group to $m=5$ further diminishes the fluctuations,
but hardly affects the tilt of what seems to
be the nearly {\it linear} divergence of the two groups. Next, for $m=6\&7,$ there
is, strangely, a discrete jump to an altogether different, but closely allied pair of trajectories.  
With $m=8(\&9),$ another jump, and so it goes with each successive
even-odd pair of multiplet selection groups.  
As $m\rightarrow N,$ the trajectories form an  
extremely well-defined V-shaped wedge, with little fluctuation at all, as we'd expect
of a mean-field limit. 
Thus we see that the dominant dynamic of secession involves, for $m\ge4,$  segregation 
into two evenly populated
opposing groups with a free interchange of individuals over the course of time. Similar {\it self-organized 
segregation} 
was reported recently by Johnson et al.,~\cite{ElFarol1,Hod} within the context of an evolutionary minority game. 
Ensemble averaging over many realizations, we have systematically studied the growth of
the population diameter over time. 
Only for triplet selection, $m=3,$ 
do we
find a fractional power-law dependence,
the diameter asymptotically scaling with an exponent very close to 3/4, our measured value being 0.74$\pm$0.01
for this one dimensional case.
 
 With the Seceder Model defined as above, the genotype space is a continuum.
Clearly, discretizing the model alters no essential features. 
Indeed, much can be gleaned by considering this discrete Seceder 
Model in its {\it deterministic} limit, wherein the most distinct individual is reproduced exactly, rather than 
yielding 
a merely approximate next of kin. One is lead to a set of nonlinear coupled ODEs~\cite{Sec1}, first-order rate 
equations for the 
concentration 
simplex
$(x_1,x_2,...,x_B),$ describing the evolution of the discrete set of $B$ genotypes possible within the population:
$\dot x_j=\sum_{i_1,...,i_m=1}^B\alpha^j_{i_1i_2...i_m}x_{i_1}x_{i_2}...x_{i_m} - x_j,$ for  $j\in (1,...,B).$
These equations transform the Seceder Model into
an evolving chemical reaction system whose dynamics are dictated by the law of mass kinetics and the 
constraints of unit dilution flux, possessing some features reminiscent
of earlier efforts on the hypercycle model, and generalized replicator equations ~\cite{eigen}.
Here,  one is looking at the stability of an $B-$branch solution generated by $m-$multiplet selection group 
dynamics.  The coefficients are zero unless the genotype/individual is the distant outlier- either
in isolation, in which case $\alpha=1$,  or as happens occasionally, sharing that distinction with

\begin{figure}
\begin{center}
  {
  \begin{turn}{270}%
    {\epsfig{file=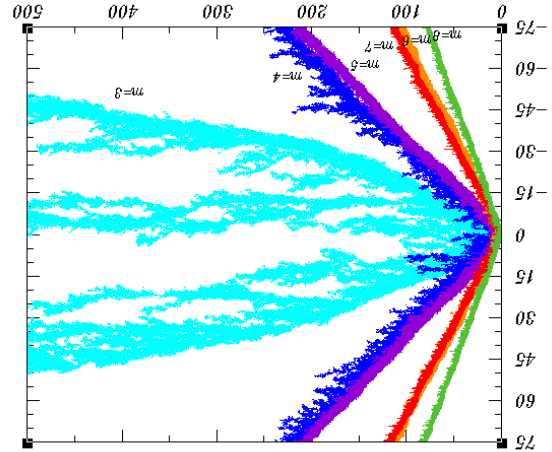,height=7.6cm,width=6.4cm} }
   \end{turn}
   }
\caption{{\it Space-time} plots of the $d=1$ stochastic Seceder Model with selection group sizes $m=3-8.$
For $m=3,$ multiple groups occur; however, for $m\ge m_c=4,$ the 
segregational dynamics yields just two repelling clusters, characterized by increasing homogeneity
in the societal Seceder limit. }
\end{center}
\end{figure}

\noindent  $p$ other 
individuals, with $\alpha=1/p$. As a practical matter, the 
coefficients, combinatoric in origin, can by generated systematically
via a multinomial expansion $(x_1+x_2+....x_B)^m$ and then carefully dividing numerical prefactors in 
appropriate 
proportions amongst relevant rate variables $\dot x_i.$ Probability conservation demands this 
connection to the multinomial expansion, but it's the parceling out of terms that guarantees the complexity 
of the model. 
With this set of ODEs in hand, essentially providing a coarse-grained 
real-space renormalization-group [RSRG] prescription of the original Seceder Model, we follow the flow equations 
for the 
concentration variables, characterizing all relevant fixed points. Within this 
broader $mB$ space, the $d=1$ Seceder Model exhibits its full richness.  As an indication of the wealth of this geometric
pattern formation, consider
for the moment triplet selection dynamic, $m=3,$ where a self-similar 
hierarchy of multibranch fixed points emerges. 
We examine, to illustrate, the case $B=4,$
for which:
$$\dot x_1= x_1^3+3x_1(x_2^2+x_3^2+x_4^2)+3x_1x_2x_3+6x_1x_3x_4-x_1$$
$$\dot x_2=x_2^3+3x_2(x_1^2+x_3^2+x_4^2)+3x_2x_3x_4 -x_2$$
\noindent Because of branch symmetry about the  central axis, the flow equations for the remaining variables 
are easily obtained via the interchange $x_1\leftrightarrow x_4$ and $x_2\leftrightarrow x_3$ and, indeed, 
the globally stable fixed point (FP) must lie within this reduced subspace, with 
mirror
variables identified. For the case at hand, invoking the constraint $x_2= 1/2-x_1$ and demanding $\dot x_1=0$ 
leads to the cubic equation $7x_1^3-6x_1^2+5/4x_1=0$, yielding
$(x_1,x_2,x_3,x_4)=({5\over 14},{1\over 7},{1\over 7},{5\over 14}),$
 as well as the less stable 2-branch solutions 
$({1\over 2},0,0,{1\over 2})$ and 
$(0,{1\over 2},{1\over 2},0). $
Of course, if we simply numerically integrate the coupled ODEs and follow the trajectories 
from a randomly generated initial condition, we flow with 100\% probability to our unique {\it superstable FP.}  
The situation for 
$B=5$ is slightly different- insisting upon symmetry $x_5=x_1$ and $x_4=x_2$, in 
addition to the normalization constraint $x_3=1-2x_2-2x_1$, we have the recurring 2-branch solutions with 
$x_1=0$ and $x_2=1/2$ and vice versa, leaving us with two coupled bilinear equations in the variables $x_1$ and 
$x_2$, represented graphically as a rotated, displaced ellipse and hyperbola within the unit square.
There are two intersection points: one full-fledged superstable 5-branch solution, $({4\over 13},{2\over 
13},{1\over 13},{2\over 
13},{4\over 13}),$ the 
other, an unstable lower-dimensional  4-branch solution  
$({1\over 4},{1\over 4},0,{1\over 4},{1\over 4}).$ Note that, in the latter instance, $x_1+x_2=1/2=x_4+x_5$, so 
that this unstable 
solution can, thanks to the gap $x_3=0$, be understood, via coarse-graining, as literally self-similar to 
its two-branch cousin $({1\over 2},0,{1\over 2}).$  This sort of hierarchical connection manifests itself regularly 
whenever 
we uncover 
a lower-dimensional fixed point; i.e., vanishing $x_i$ in the branch structure. The most 
compelling instance of this phenomenon appears when we search for a superstable  8-branch FP. In fact, there 
is none.  The ODE flows converge on a peculiar 6-branch solution, $({11\over 40},{1\over 8},0,{1\over 10},{1\over 
10},0
,{1\over 8},{11\over 40}),$ which is {\it exactly} 
self-similar to the strongly attractive 3-branch fixed point $({2\over 5},{1\over 5},{2\over 5}).$ Interestingly, 
this {\it lesser} 
8-branch 
solution is distinct from the straight out 6-branch, roughly (0.30,0.12,0.08,0.08,0.12,0.30), easily 
shown to be irrational,
as is the 7-branch and all those beyond 8. The 9, 10 and 12-branch FPs show no zeros, but such behavior 
becomes
increasingly rare. The 11-branch solution has two gaps, $x_4=x_8=0,$ but with $x_1=x_{11}\approx0.271, 
x_2=x_{10}\approx0.089, x_3=x_9\approx0.044$ and $x_5=x_7\approx0.047, x_6\approx0.100,$ can be 
coarse-grained to a broad 3-branch again, though in this case only approximately, self-similar to our 
dominant FP $({2\over 5},{1\over 5},{2\over 5});$ likewise, the 15-branch, although 19 and 21-branch FPs show three 
gaps and a 
self-similarity to the 4-branch $({5\over 14},{1\over 7},{1\over 7},{5\over 14}).$  

An additional payoff of this RSRG treatment of the Seceder Model is an explanation of 
the relative stability of 2 and 3-branch solutions for triplet ($m=3$) and higher multiplet 
($m\ge4$) selection groups. Recall Figure 1, which made it clear that for large $m,$ the dynamics of the 
$d=1$ stochastic Seceder Model are controlled entirely by the strongly attractive fixed point of the 2-branch 
pattern.  For triplet selection, however, a hierarchy of multi-branch solutions is manifest, 
which meant the frequent appearance of a 3-branches, and somewhat
occasionally 4 and 5-branches, but a complete absence of 2-branch behavior.  The essential dichotomy can be 
understood
graphically by following the flows for 3-branch dynamics in the 
deterministic Seceder Model, assuming triplet selection group $m=3,$  illustrated in Figure 2a, where we show the 
[111] plane 
$x_1+x_2+x_3=1.$ We find our superstable 
fixed point, $({2\over 5},{1\over 5},{2\over 5}),$ within
the unit triangle. We note, in particular, that the outmost 2-branch solution  $({1\over 2},0,{1\over 2})$ is {\it 
unstable}
to small perturbations off the edge.  In turn, the single branch FPs at the triangle vertices are 
entirely unstable.  As $m\rightarrow 4,$ however, this interior FP merges with that at the midpoint on the 
triangle's lower edge, 
reversing the flow and stabilizing the 2-branch dynamic.
From this vantage point, it is clear that quartet, rather than triplet, selection is the marginal case, a fact   
quickly 
confirmed by a stability analysis of the $({1\over 2},0,{1\over 2})$

\begin{figure}
\begin{center}
  {
  \begin{turn}{0}%
    {\epsfig{file=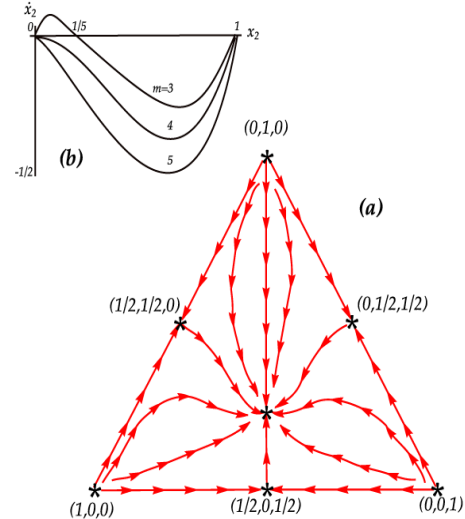,height=7.0cm,width=6.0cm} }
   \end{turn}
   }
\caption{a) RSRG flows in $x_1x_2x_3-$space for the $d=1$ deterministic Seceder Model. A superstable 3-branch
FP exists within the equilateral triangle for $m<4$ only. b) Middle branch growth rate, $\dot x_2,$ for different multiplet 
selection group sizes, $m=3-5,$ showing stability of the 2-branch solution (1/2,0,1/2) for quartet selection and
greater.}
\end{center}
\end{figure}

\noindent     FP for arbitrary $m.$  To linear order, we find
$\dot x_2 = \bigl[2{m\choose1}\bigl({1\over2}\bigr)^{m-1}-1\bigr]x_2 + O(x_2^2),$ so we flow back to vanishing 
$x_2$ for 
$2m\le2^{m-1};$ i.e., $m\ge 4,$ since prefactor of the quadratic term is negative for the marginal value $m_c=4:$ 
see Figure 2b, 
which shows the full behavior. 
In fact, for a continuously variable selection group of size $m=4-\varepsilon,$ the {\it perturbatively 
stable} 3-branch FP is located off the triangle's lower edge at
$x_2={\varepsilon\over12}(\ln 2-{1\over4}).$  

As one considers 
the stability of higher ($B>3$) multibranch FPs, $m_c=4$ remains the {\it upper critical
size} of the selection group at, and above which, mean-field deterministic pattern formation holds sway.  For 
example, with 
quartet selection in the context of the 4-branch solution (i.e., $m=4,B=4$), the superstable FP lies at the 
midpoint of the edge 
connecting 
the $x_1$ and $x_4$ vertices of the unit tetrahedron- that is, $(x_1,x_2,x_3,x_4)=({1\over 2},0,0,{1\over 2}).$ All 
initial starting 
points
within the tetrahedron flow outwards to the FP on this edge. While the other five edges of the tetrahedron are 
stable along
their lengths, they're unstable in all other directions; the vertices, corresponding to 1-branch solutions, are 
maximally
unstable. For triplet selection, the superstable 4-branch FP is, as mentioned earlier, $({5\over 14},{1\over 
7},{1\over 7},{5\over 
14}),$
 which by contrast lies 
within the 
tetrahedron.  There are also stable 3-branch FPs in this case located in the faces of the tetrahedron, where one of 
the 
$x_i=0,$
but these are unstable to perturbations toward interior of the tetrahedron. Vertex FPs are unstable to edge FPs, 
which in turn are 
unstable
to face FPs, etc.  Interestingly, these  findings might suggest, at least initially, that the greatest stability is 
associated with the largest number of branches;
i.e, $(B-1)-$branch being unstable to $B-$branch solutions, etc.  However, the absence of superstable 8-branch FP, 
for starters, and  
the subsequent appearance of gaps for $B=11,13,17$ in the spectrum indicate that any runaway tendency toward 
proliferation
of 
branches
from tip splitting will be cut off.

\begin{figure}
\begin{center}
  {
  \begin{turn}{270}%
    {\epsfig{file=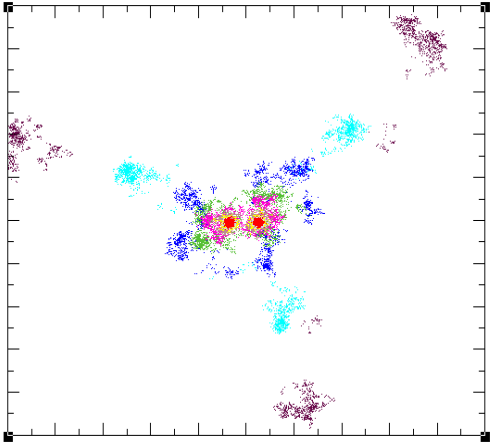,height=7.0cm,width=6.4cm} }
   \end{turn}
   }
\caption{ Despite greatly biased initial conditions, with two distinct localized groups, the $d=2$ triplet Seceder 
model evolves the population toward an endpoint of three equidistant, separating clusters. 
Here, $N$=2000, 
and the temporal snapshots, indicated by different colors, correspond to successive times $t=2^i,$ for $i=2-8;$
red$-$mauve, respectively.
 }
\end{center}
\end{figure}

\noindent  
   Indeed, that is precisely the characteristic behavior of the triplet Seceder Model, where 
3-branch
dynamics are typically seen, with occasional 4, 5 or 6-branch runs. 

We should stress, in this regard, that the 3-branch solution
is an extraordinarily robust feature of this model, becoming
even more so in higher dimensions, where the genotype is specified 
by an $d-$component vector rather than a single real number, the case we've focussed on thus far. For example, in 
$d=2$, an
initially homogeneous, or highly polarized population for that matter- see Figure 3, will eventually segregate into 
three distinct 
groups
heading off along the symmetry axes of a triangle, each cluster equidistant from the other two. In $d=3$ 
dimensions, 
we might expect four groups, perhaps, localized at the corners of an expanding tetrahedron, preserving the notion 
of 
equal distance. Interestingly,  however, this does not happen at all. Again, 
we observe the formation of just three
groups- note Figure 4; the effect is stronger still for $d\ge 4.$ Apparently, asymptotic higher-dimensional 
secession involves
 segregational collapse to a greatly reduced, two-dimensional, subspace- the hyperplane defined by three fuzzy 
points of 
an expanding 
 equilateral triangle whose angular orientation may vary from one realization to the next, but whose essential
 geometry does not.  Interestingly, this dimensional reduction can be understood within the context
  of the deterministic model- one considers the stability of $d+1$ equally separated groups in $d$ dimensions to 
perturbations
  (ultimately, statistical in nature) that bring one group closer to the rest~\cite{flo}. For $d\ge3,$ we find that 
the minority group is unstable and will go extinct, whereas for $d<3,$ the zero FP associated with this vanishing 
group reverses stability, yielding three separatist clusters whose relative population sizes are set by the degree 
of symmetry breaking. 
  
  In sum, we have revealed the mean-field limit of the multidimensional Seceder Model. For selection multiplet
   sizes $m\ge 
 m_c=4,$ the nonequilibrium  dynamics produce
   
\begin{figure}
\begin{center}
  {
  \begin{turn}{0}%
    {\epsfig{file=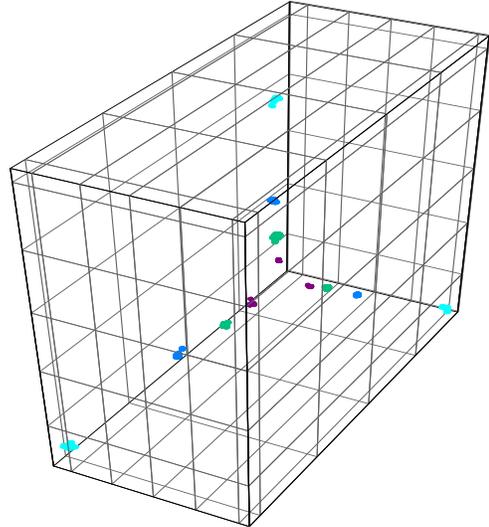,height=7.0cm,width=6.4cm} }
   \end{turn}
   }
\caption{Self-organized secession in $d=3,$ as well as higher dimensions, collapses to the plane defined by 
three divergent groups. Population size $N$=1000; generations $t=2^{9-12}.$}
\end{center}
\end{figure}

  \noindent   a steady-state with two opposing groups, independent of $d.$ In the extreme societal Seceder 
limit
 ($m\rightarrow N$), the noisy dynamics dies away, leaving two tightly knit groups.  For $m=3,$ 
 multiple groups are typical, with 
 three the norm. Higher dimensional genotypes/strategies 
 produce, surprisingly, no further fragmentation. 
Using a coarse-grained RG 
 prescription, which discretizes and renders deterministic the model, we analytically uncover a self-similar 
hierarchy of multiple
 branch FPs in a gapped spectrum.  Additional work, concerning the intermittent extinction dynamics of 
individual 
groups,
 early-time transient behaviors, as well as kinetic symmetry-breaking phenomena, will be 
reported 
elsewhere~\cite{Tim}.     

Financial support for Tim HH has 
been provided by NSF DMR-0083204, Condensed Matter Theory.


\begin{thebibliography}{99}
\bibitem{flo} we break the symmetry by hand for the odd group, writing its ODE coefficient
	$\alpha=(1-K)/d.$  Stability analysis reveals that critical symmetry-breaking parameter $K_c=2/(d-1)$ 
decreases with dimensionality; statistical fluctuations thereby induce cascadal reduction to $d=2.$
\end{thebibliography}
\end{document}